\documentstyle[preprint,prd,aps,eqsecnum,epsfig,amssymb]{revtex}
\tightenlines
\everymath={\displaystyle}

\begin{document}

\title{Three-body confinement force in a realistic 
constituent quark model}

\author{Z. Papp \footnote{e-mail address: zpapp@csulb.edu}}
\address{Institute of Nuclear Research, H-4001 Debrecen, Pf. 51, 
Hungary}
\address{Department of Physics and Astronomy,
California State University, Long Beach, CA-90840, USA}

\author{Fl. Stancu \footnote{e-mail address: fstancu@ulg.ac.be}}
\address{University of Li\`ege, Institute of Physics B5, Sart Tilman,
B-4000 Li\`ege 1, Belgium}

\date{\today}
\maketitle

\begin{abstract}
We show that in realistic constituent quark models based on a 
two-body confinement interaction color states appear in the
middle of the experimentally known spectrum. To avoid this situation
we implement a three-body confinement interaction, introduced
on an algebraic basis, into a semirelativistic version of 
the Goldstone boson exchange constituent quark model and 
by solving the Faddeev equations we show that this interaction 
can increase the gap between singlet and color states,
such as the latter can be ignored and the known baryons 
can be described as simple $q^3$ systems. We analyze the effect 
of a $\Delta$- and a Y-shape three-body interaction.
\end{abstract}

\vspace{0.2cm}
\begin{flushleft}
\end{flushleft}
\vspace{0.2cm}


\section{Introduction}

In this study we are concerned with constituent quark models where the 
confinement interaction contains the two-body color operator  $F_i \cdot F_j$
\cite{GL81}. This means to assume that the one-gluon exchange picture
survives to all distances. There is no real justification for applying
this rule to the confinement, except simplicity. This relatively easy model 
has been widely used in baryon spectroscopy together with a
hyperfine interaction generated either by the one-gluon exchange 
\cite{RGG} or by the Goldstone boson (pseudoscalar meson) exchange
\cite{GR96}. The  $F_i \cdot F_j$ model has often been revived as being a 
convenient framework for studying $q^2 {\overline q}^2$ systems (tetraquarks),
see e. g. Ref. \cite{WI}, $q^4 {\overline q}$ systems (pentaquarks), 
see e. g. Ref.  \cite{penta} 
or $q^6$ system (the nucleon-nucleon problem), see e.g. Ref. \cite{HARVEY}.  

The present work is devoted to baryons, seen as $q^3$ systems. The starting
point is the observation that the presence of the  $F_i \cdot F_j$
color operator implies that the Hamiltonian, besides color singlets,
also possesses
color octet and decuplet states at low energies. These states are undesirable. 
Their existence is in conflict with the experiment and
with conclusions drawn from lattice or continuous formulations of QCD in
the confining phase.  
In practice, these states have tacitly 
been ignored,
relying on the assumption that they do not exist.
However, as shown below, colored states are found close to singlets in the 
spectrum of consituent quark models.
Therefore they should either be shifted to higher energies, 
by a change in the dynamics 
of the quark model,  so that they could be ignored,
or otherwise transformed into colorless states  by coupling 
to gluons as for example in \cite{LIPKIN,CHENG_LI,LINDE},
or to color octet sea $q \overline q$ pairs   
as for example in \cite{HELMINEN}.

A solution in the first direction 
has recently been proposed by Dmitrasinovic \cite{DMI}. 
Although OCD has a local exact SU$_{C}$(3) symmetry, in 
QCD-inspired potential models this symmetry appears as global, by construction.
The observation was that by considering only two-body confining forces 
in a Hamiltonian model implies that the global SU$_{C}$(3) symmetry is   
implemented only in a restricted way, namely 
the Hamiltonian is expressed in terms of 
the quadratic (Casimir) invariant operator of SU(3) only. More generally,
one could  express any quark model Hamiltonian inspired by QCD  
in terms of every SU(3) invariant operator.
Based on this argument, in Ref. \cite{DMI}  a three-quark 
confining potential
that depends on the cubic invariant operator of SU(3) has 
been added to the usual two-body confinement. In a qualitative way, 
it was shown that such an interaction can 
increase the gap between the singlet and color states, provided
its strength has a specific sign and range.
Besides its implications on the spectrum of ordinary $q^3$ systems, its 
role in  exotic $q^2
\overline {q}^2$ hadrons has also been considered. 

The three-body confinement interaction of Ref.\ \cite{DMI} has  
subsequently been analysed in Ref.\ \cite{PS} in the context of a harmonic 
oscillator confinement where the size of its effect 
on the gap between the 
color and singlet states in the spectrum of a $q^3$ system has 
been estimated. 
Also the $q^6$ system with relevance to 
the nucleon-nucleon problem has been discussed.

Below  we first show explicitly how much the 
singlet and color states of a realistic constituent quark model 
appear close to each other. Next, by introducing a three-body
confinement force related to the cubic invariant of SU(3), 
we demonstrate that one can separate them largely enough so that the 
color states can be ignored. For this purpose, here   
we consider the semirelativistic 
constituent quark model of Ref.\ \cite{FRASCATI},
into which we implement a three-body confinement interaction. 
By using the Faddeev approach of Ref.\ \cite{PAPP} we calculate the lowest singlet, octet and 
decuplet states
of a three-quark system, 
with this additional three-body confinement force. 

\section{Basis states}
Here we present the nucleon sector, namely S=1/2 and I=1/2. 
The lowest color octet and decuplet $q^3$ states compatible 
with the Pauli principle are displayed in Tables \ref{octet}
and \ref{decuplet} respectively.
For orientation, their excitation energy $E_N-E_0$,
estimated in a two-body harmonic oscillator confinement,
is shown in the last column.
In the same harmonic oscillator confinement the first colorless  
negative parity state is 395 MeV above the ground state.
Therefore one can see that the $0^+$ color octet and decuplet states 
appear substantially below this negative parity state.
In the following we shall consider only the lowest color states,
namely the $0^+$ and $1^-$ octets and the  $0^+$ decuplet.
It is necessary
to displace all color states at energies where they can safely be neglected,
if the nucleon resonances are to be interpreted as pure $q^3$ excitations.
This can indeed be achieved by an
additional three-body force, as shown below.

\section{The Hamiltonian}

The realistic total Hamiltonian presently under discussion  
has the form
\begin{equation}\label{HAMILT}
H = \sum\limits_{i} \left(m_i^2 + p_i^2\right)^{1/2}
 + V_{2b} + V_{3b} + V_{\chi}~,
\end{equation}
where $\vec p_i$ and $m_i$ are the momentum and the constituent mass of the
quark $i$,
$V_{2b}$ the two-body confinement, 
$V_{3b}$ the three-body confinement and $V_{\chi}$ the hyperfine
interaction.  
Taking $V_{3b}$ = 0 one recovers the chiral constituent quark model 
of Ref.\ \cite{FRASCATI} which is the simplest semirelativistic version
(no tensor, no spin-orbit)
of the model originally proposed in Ref.\ \cite{GR96}, where the 
hyperfine interaction 
is spin and flavor dependent. 
The origin of such an interaction is thought to be the pseudoscalar meson 
exchange between quarks. This model,
either in a nonrelativistic or a semirelativistic form,
reproduces well the spectra of light baryons
and in particular the correct order of positive and negative parity 
states. 
It has been shown that the addition of a tensor interaction to
the spin-spin interaction in the Goldstone boson exchange model
has a minor effect 
and thus maintains the quality of the 
spectrum \cite{TRIESTE}. Therefore, in the present study where the
emphasis is made on the confining interaction the tensor force can be ignored. 
Although we have chosen a particular model,  
we expect our conclusions to be relevant  
for any reasonable constituent quark model with three valence quarks.
The detailed parametrization of $V_{\chi}$ can be found in 
Ref.\ \cite{FRASCATI} and we do not reproduce it here, 
the main emphasis
being laid on the confinement potential. We use 
a $V_{2b}$ term of the form
\begin{equation}\label{V2b}
V_{2b} = \sum\limits_{i<j}{\mathit v}^c (r_{ij})~(\frac{7}{3} 
+ F^{a}_{i} F^{a}_{j})~,
\end{equation}
where $F^{a}_{i} = \frac{1}{2} \lambda^{a}_{i} $ is the color charge
operator of the quark $i$, $\lambda^a$ are the Gell-Mann matrices 
and 
\begin{equation}\label{RADIAL}
{\mathit v}^c (r_{ij}) = V_0 + \gamma r_{ij}~, 
\end{equation}
with $\gamma=2.33$ fm$^{-2}$
and $V_0= -416$ MeV as in Ref.\ \cite{FRASCATI}. The color part
of $V_{2b}$ is 
consistent with Refs.\ \cite{DMI,PS} but different from that of \cite{FRASCATI}.
It stems from the stability condition of a $q \overline q$ pair
but it depends on an arbitrary constant which is fixed to 7/3.
Anyhow, we rescale this interaction such as to reproduce the spectrum
obtained in \cite{FRASCATI} (see below).

The three-body confinement interaction has the form \cite{DMI} 
\begin{equation}\label{3BTOTAL}
V_{3b}=V_{ijk} = {\mathcal V}_{ijk} {\mathcal C}_{ijk}~,
\end{equation}
where ${\mathcal V}_{ijk}$ is the radial part and ${\mathcal C}_{ijk}$
is the color operator
\begin{equation}\label{3B}
{\mathcal C}_{ijk} = d^{abc}~F^{a}_{i} ~F^{b}_{j}~ F^{c}_{k}~,
\end{equation}
with $d^{abc}$ some real constants,
symmetric under any permutation of indices and
defined by the anticommutator of the Gell-Mann matrices  as
$\{ \lambda^a,  \lambda^b \} = 2 d^{abc}~\lambda^{c}$ .
The operator (\ref{3B}) can be rewritten in terms of the two
independent invariant operators of SU(3) as 
\begin{equation}\label{EQ4}
{\mathcal C}_{ijk} =
\frac{1}{6}~ [~  C^{(3)}_{i+j+k} - \frac{5}{2} C^{(2)}_{i+j+k} +
\frac{20}{3}~ ]~,
\end{equation}
where $C^{(2)}$ is the quadratic (Casimir) and $C^{(3)}$
the cubic invariant. The expectation values of (\ref{EQ4})
can thus easily be obtained from the eigenvalue of $C^{(2)}$ and
$C^{(3)}$ \cite{book} for any irreducible representation $(\lambda \mu)$
of SU(3). These eigenvalues are
\begin{equation}
\langle C^{(2)} \rangle = \frac{1}{3} (\lambda^2 + \mu^2 + \lambda \mu
+ 3 \lambda + 3 \mu )
\end{equation}
and  
\begin{equation}\label{AUTOG}
\langle C^{(3)} \rangle = \frac{1}{18} (\lambda - \mu)
(2 \lambda + \mu + 3)(\lambda + 2 \mu + 3)~.
\end{equation}

Our attempt to choose a form for ${\mathcal V}_{ijk}$ is related to the present 
knowledge of confinement.  
Lattice calculations are ambiguous about the static three-quark 
potentials in baryons.
Both Y-shapes \cite{TAKAHASHI} and $\Delta$ shapes \cite{BALI,ALEXANDROU}
are supported. 
The difficulty to distinguish between them is due to the fact that
the difference between the two gives at most 15 \%  in the dominant
area law for the baryonic potential 
(for a recent review see
e. g. \cite{BALI2}). 
Note that these results concern color singlet states only while here
we deal with both color singlet and octet states.
Potentials between static color sources in higher representations of SU(3),
including the octet one, do also exist \cite{BALI3}. They satisfy the 
so-called Casimir scaling which means that potentials between sources in
different representations are proportional to each other such as their 
ratios is given by the respective ratios of the eigenvalues of the
corresponding (quadratic) Casimir operators.  On the other hand the
scaling proportional to the number of fundamental flux tubes
embedded into a higher representation coincides with the Casimir
scaling.
In this situation we deemed reasonable to consider the working Ansatz
that ${\mathcal V}_{ijk}$  
has either a triangular shape 
\begin{equation}\label{DELTA}
 {\mathcal V}_{ijk} = {\gamma} ~c~[|\vec{r_i}-\vec{r_j}|
+|\vec{r_j}-\vec{r_k}|+|\vec{r_k}-\vec{r_i}|]~,
\end{equation}
or a Y-shape  
\begin{equation}\label{Y}
 {\mathcal V}_{ijk} = {\gamma}~c~[|\vec{r_i}-\vec{r_0}|
+|\vec{r_j}-\vec{r_0}|+|\vec{r_k}-\vec{r_0}|]~,
\end{equation}
with $\vec{r}_0$ the point
where the three flux tubes meet such as to satisfy the SU$_{C}$(3) gauge
invariance (see e.\ g.\ \cite{CKP}).
Taking $c = 1$ and $\gamma = 1/2 \sqrt{\sigma}$
with a strength tension  $\sqrt{\sigma} \approx 1 $ GeV fm$^{-1}$ 
in (\ref{DELTA}), 
one recovers the linear confinement potential used in 
realistic consituent quark models as for example the model
of Ref. \cite{FRASCATI} where the form (\ref{DELTA})
is used in conjunction with a two-body color operator
$F_i \cdot F_j$ which brings the factor 1/2 in $\gamma$.
Taking $c = 1$ and $\gamma =  \sqrt{\sigma}$ in (\ref{Y})
one recovers the confinement potential used for example in Refs. \cite{CKP},
\cite{SS} or \cite{CI} where no color operator is present
in the confinement part, thus one deals with colour singlets only.
Here we use the radial form  (\ref{DELTA}) or (\ref{Y})
in  Eq. (\ref{3BTOTAL}), i.e. in conjunction with a three-body color operator.  
In our calculations  
the value of $\gamma$ appearing both in (\ref{DELTA})  and (\ref{Y})
is taken to be the same as in Eq.\ (\ref{RADIAL})
and the parameter $c$ thus represents the relative strength of the 
three-body versus the two-body force. By assuming the same triangular shape
both in $V_{2b}$ and $V_{3b}$ 
analytic calculations
can be carried for a while. In this way
in Ref.\ \cite{DMI} it was found that $c$ must be located in the interval  
$-\frac{3}{2}~<~c~<~\frac{2}{5}.$
The upper limit ensures that the lowest
color singlet is below the lowest color octet state. The lower limit was   
required by the stability condition of the nucleon, 
$\langle V_{2b} + V_{3b} \rangle > 0$. 
In Ref.\ \cite{PS} some arbitrariness was noticed for 
the lower bound because this is related to the choice of
the color operator in (\ref{V2b}). 
However, as the numerical results of Sec. IV will reveal,
the above range of $c$
is entirely satisfactory for our discussion of the triangular shape, because
the gap of between the colorless and color states will turn out to be large
enough.
In fact the common conclusion of 
\cite{DMI} and \cite{PS} was that $c$ must be negative in order
to obtain an increase in the gap between the color octet and singlet states
in a $q^3$ system.
Therefore 
in the calculations below related to the triangular shape, we take 
\begin{equation}\label{INEQ1}
-\frac{3}{2}~<~c~<~0~.
\end{equation}

The Hamiltonian (\ref{HAMILT}) was solved by using the Faddeev approach of
Ref.\ \cite{PAPP}, adequate for confining potentials. 
The necessary expectation values for the two-body color
operator appearing in (\ref{V2b}) are given in Table \ref{twobody}. The
expectation values of the three-body color operator (\ref{EQ4}) are taken
from \cite{DMI} or \cite{PS}. These are 10/9, -5/36 and 1/9 for
the singlet, octet and decuplet SU(3) states, respectively.

A particular advantage of the Faddeev approach is that the incorporation
of permutation symmetry is very easy. For identical quarks, as in this
case, the three Faddeev components of the three-quark
wave function have the same functional form in their 
own Jacobi coordinate systems.
Therefore the three equations can be reduced to a single one. From the 
structure  
of this equation it follows that the correct symmetry of the wave function under
the exchange of any two particles is automatically guaranteed if the correct
symmetry is implemented in one of the three components.
In our calculation a bipolar harmonic basis was used,
combined with spin, isospin and color basis states.
If we select the basis states such as
\begin{equation} \label{permut}
(-)^{l + s + i + c} = - 1~,
\end{equation}
where $l$, $s$, $i$ and $c$ are the relative angular momentum, spin, isospin
and color quantum numbers of a two-quark pair, the Pauli principle for the 
three-quark system is satisfied.
We have $(-)^{c}$ = 1 when $[\tilde{f}]_C=[2]$ and $(-)^{c}$ = - 1 when 
$[\tilde{f}]_C=[11]$, where $[\tilde{f}]_C$ is a given partition in the 
color space.
If we denote the total angular momentum  by $L$ and the 
total parity by $P = (-)^{l + \lambda}$, where $\lambda$ is the relative
angular momentum of the third particle with respect to the pair,
then the lowest color octets must have $L^P = 0^+, 1^{-}$ 
and the lowest decuplet $L^P = 0^+$.
In Table \ref{pauli} we 
show the structure of 
the lowest color states with the corresponding 
quantum numbers of their components. 
This treatment of permutation symmetry as well as the whole numerical procedure
were checked against the exact harmonic oscillator results of Ref.\ \cite{PS}.


\section{Results}

As mentioned above, our purpose is to understand the implications of 
a three-body color confinement interaction in a realistic model.  We look
separately at its effect either due to the $\Delta$-shape (\ref{DELTA}),
or to the Y-shape (\ref{Y}).  


\vspace{1cm}
\subsection{The $\Delta$-shape.}
This shape is currently used in conjunction
with a two-body $F_i \cdot F_j$ color operator as an approximation
to the Y-shape  
\cite{CKP}. Here we use it both in $V_{2b}$ and $V_{3b}$ 
as discussed above. Then the contribution of the color part of 
$V_{2b}$ and $V_{3b}$ sum up together to $\chi_i$ (i=1,8,10)
as in Eq.\ (14) of Ref.\ \cite{PS}.
\begin{equation}\label{CHI}
{\chi_i} = \left\{ \renewcommand{\arraystretch}{2}
\begin{array}{cl}
 \frac{5}{3} + \frac{10}{9} c   &\hspace{1.1cm} \mbox{ i=1 (singlet)} \\
 \frac{13}{6} - \frac{5}{36} c
 & \hspace{1.1cm} \mbox{ i=8 (octet)} \\
 \frac{8}{3} + \frac{1}{9} c
  & \hspace{1.1cm} \mbox{ i=10 (decuplet)}
\end{array} \right. 
\end{equation}
 
In Fig.\ 1 we show the dependence of some eigenvalues of the
Hamiltonian (\ref{HAMILT}) as a function of $-c$ for $-1.4~ \leq c~ \leq 0$.
The case $c$ = 0 is the model \cite{FRASCATI} for
which we reproduced the ground state nucleon mass $m_N$=940 MeV, 
and the resonance masses N(1440)1/2$^{+}$= 1459 MeV,
N(1535)1/2$^{-}$ = 1522 MeV 
and N(1710)1/2$^{+}$= 1783 MeV. 
For $c=0$
the lowest color octets $0^+$ and $1^{-}$ 
acquire the masses 1536 MeV and 
1758 MeV respectively and the $0^+$ color decuplet has 2077 MeV. 
Thus in a 
constituent quark model with three valence quarks the color states
are so low that they cannot be ignored. If the three-body interaction
(\ref{DELTA}) is switched on, 
the singlet states go substantially down and the color states
go slightly up (or down) and a clear gap between the lowest color
singlet S=1/2, I=1/2 states and the color $q^3$ states
emerges. This gap increases when
$c$ is decreased. At 
$c~ \sim -1.4$ it aquires a substantial value of more than 1500 MeV. 
  
Note however that the singlet states
part of the spectrum shrinks and one has to bring it back close to
the experimental spectrum. This can be achieved by a readjustment of 
the model parametres defined in Ref. \cite{FRASCATI}. In particular,
in order to enlarge the distance between color 
singlet states one has to increase the
strength tension $\gamma$ in (\ref{RADIAL}) which will amount to
an increase of the gap between the color and colorless states as well.
It is then inferred that the gap will be large enough so that 
color states can safely be neglected.

Before ending this subsection
a word should be mentioned about the influence of the choice
of the additional constant 7/3 in (\ref{V2b}) on the results.
As shown by Dmitrasinovic \cite{DMI} the presence of this positive
constant is crucial for the stability of a $q {\overline q}$ pair.
In Ref. \cite{PS} some arbitrariness was noticed regarding this 
constant.  In the $q^3$ spectrum, a change in the above
additional constant amounts to a shift of the whole spectrum
and some change in the slope of the confinement. 
If one wants to restore the spectrum to its initial form one has to
change the arbitrary constant $V_0$ and the quantity $\gamma$ in
Eq. (\ref{RADIAL}). 
For example, if the additional constant is increased,
$\gamma$ must be decreased. As a consequence the gap between the color and 
singlet states  
decreases.  But this can be compensated 
by further decreasing the lower limit of $c$
below - 1.5, which is allowed in the way discussed  
in \cite{PS}.

\vspace{1cm}
\subsection{The Y-shape.} 
This is a genuine three-body force, both in the
coordinate and the color space as well.  
Then the color part contributions of $V_{2b}$
and $V_{3b}$ does not sum up analytically to $\chi_i$ as in Eq. (\ref{CHI}),
because the expectation value of the radial part of $V_{2b}$ is different 
from that of $V_{3b}$. Hence the numerical calculation of the entire spectrum 
is slightly more complicated.

In the Y-shape, the flux tubes meet at 120$^0$ in order to ensure the 
minimum energy. If one of the interior angles of the flux-tube 
configuration is greater than 120$^0$ the minimum energy condition
cannot be satisfied and the corresponding flux-tube collapses to 
a point which means that $\vec{r}_i$= $\vec{r}_0$ for $i$ = 1, 2 or 3.  
Thus the Y shape moves continuously into a two-legged flux-tube
configuration, where the legs meet at an angle greater than 120$^0$.
The geometrical arguments are clearly given in Ref. \cite{CKP} for example.
 
\begin{figure}
\begin{center}
\psfig{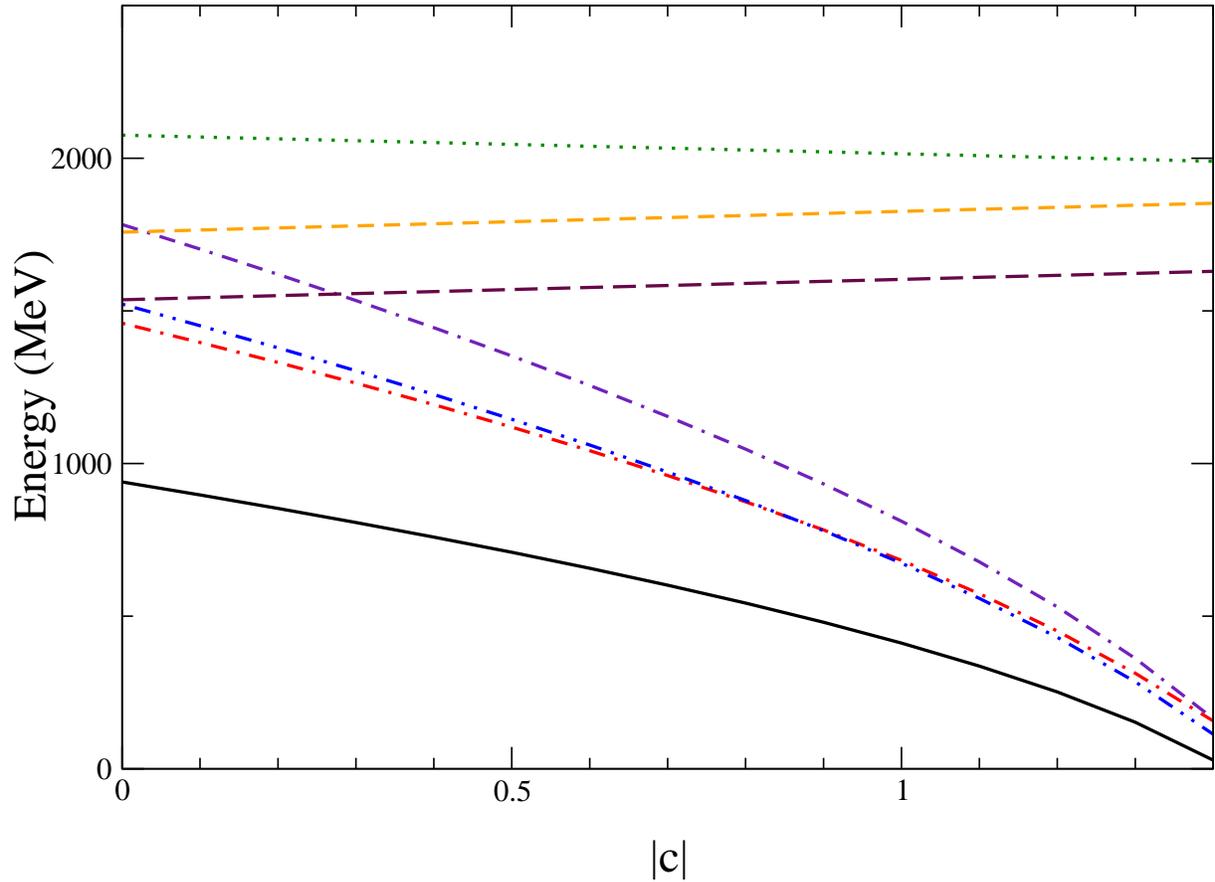}
\end{center}
\caption{\label{Fig1}
S=1/2, I=1/2 eigenvalues 
of the Hamiltonian (\ref{HAMILT}) with the $\Delta$-shape
three-body interaction  (\ref{DELTA})
as a function of $-c$. The solid 
line is the ground state nucleon, the dash-dotted line the N(1440) 
resonance (Roper), the double-dot-dashed line the lowest negative parity
state N(1535), the double-dash-dotted line 
the N(1710) resonance.
The color octets $0^+$ and $1^{-}$ are represented by 
long- and short-dashed lines respectively and the $0^+$ color
decuplet by a dotted line. }
\end{figure}

\begin{figure}
\begin{center}
\psfig{figure=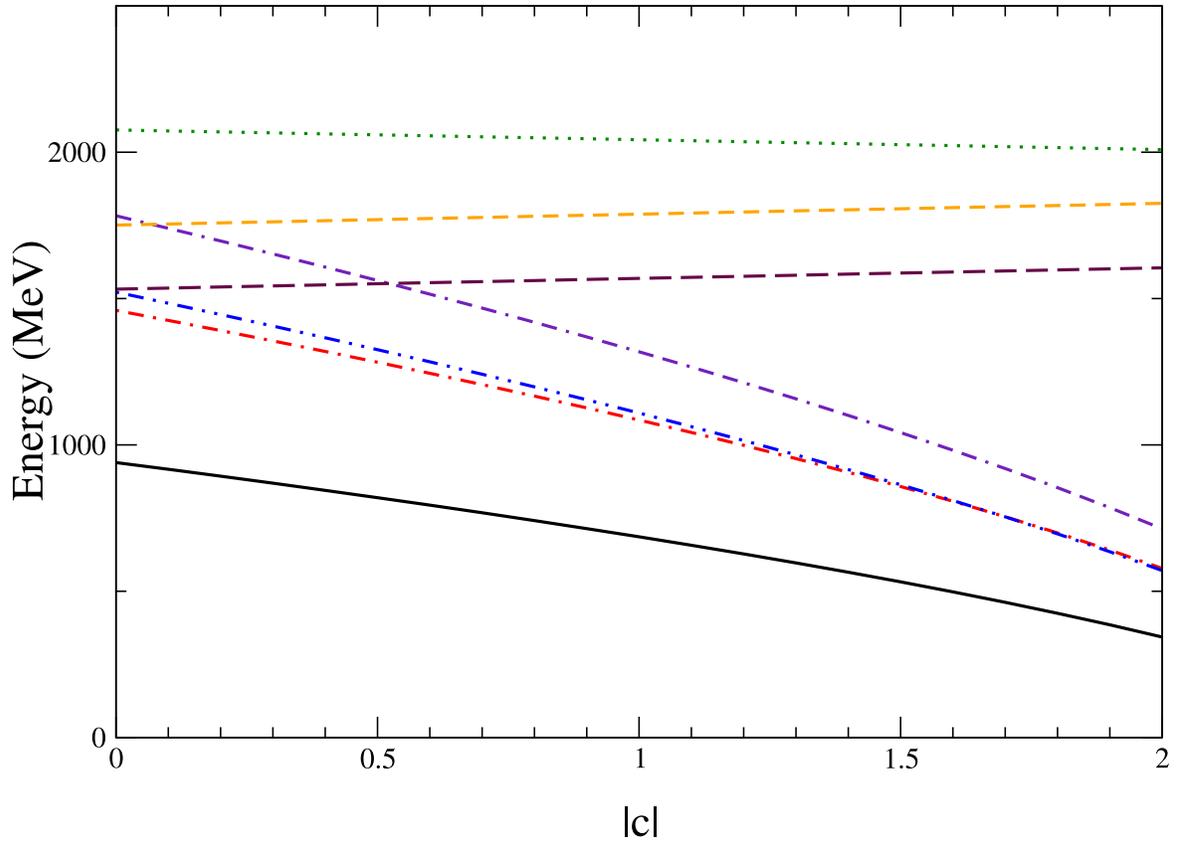, width=18cm}
\end{center}
\caption{\label{Fig2}
Same as Fig.~\ref{Fig1} but for the Y-shape interaction
(\ref{Y}). }
\end{figure}

The spectrum associated to the Y-shape (\ref{Y}) is presented in 
Fig.~\ref{Fig2}. In this case the lower limit on $c$ 
imposed by the inequality (\ref{INEQ1}) is no more valid but $c$  
must remain negative. So we varied the parameter $c$ between zero and - 2.
At $c = 0$ the spectrum is the same as in Fig.~\ref{Fig1}.

The trend is similar to the $\Delta$-shape case, but the mass of every
colorless state vary more slowly with the strength 
$|c|$ of the three-body Y-shape force. 
If one would rescale the $|c|$-axis by a factor 1/2, the $\Delta$ and the 
Y-shape results would look closer to each other.

\begin{figure}
\begin{center}
\psfig{figure=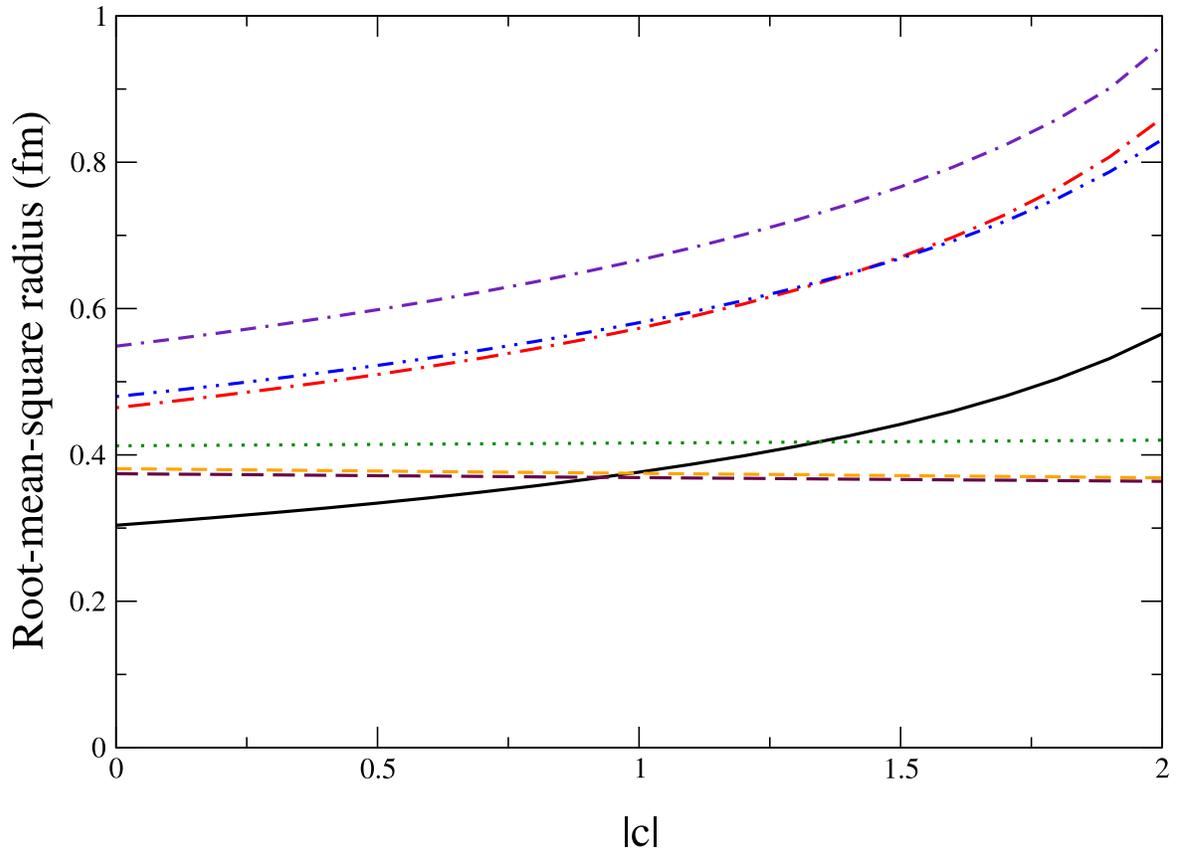, width=18cm}
\end{center}
\caption{\label{Fig3}
Root mean square radii of states of  Fig.~\ref{Fig2}.}
\end{figure}

   To better understand the role of the three-body confinement force we 
also calculated the root-mean-square radii associated to the states
of Fig.~\ref{Fig2}. These are displayed in Fig.~\ref{Fig3}.
One can see that the quark core 
radii of the ground state nucleon and its lowest 
N(J=1/2)  
resonances increase slowly with $|c|$.
This 
is obviously the effect 
of the decrease of the confinement contribution through the
addition of the three-body term.
On the other hand, the radii of the color
states  remain practicaly unchanged, similar to their energies.

\section{Conclusions}

Through the example of the Goldstone boson exchange 
model\ \cite{FRASCATI} 
we have shown that the spectrum of a realistic constituent quark 
model with a pairwise color  confinement operator
and a linearly increasing behaviour
can accomodate both singlet and color low-lying $q^3$ states.
The lowest octet and decuplet
color states appear in the middle of the observed spectrum
which means they cannot be ignored. In this situation there are
two alternatives:

1) To keep simplicity. Then, as shown here,
the addition of a three-body confinement interaction
can increase the gap between 
singlet and color states by 1. 5 GeV or more.  
Then the color states can be neglected
in calculations and the presently observed baryons can be described as pure 
$q^3$ states.
To restore the quality of the spectrum one has to 
refit the parameters of the model, including the strength tension
$\gamma$, which has to be increased in order to enlarge the spacing
between colorless states. The increase of $\gamma$ will lead to a further
increase of the gap between the colorless and color states.

2) If however the color states are maintained low in the spectrum
of a Hamiltonian with a pairwise confinement interaction only,
one has to give up the simple $q^3$ picture of baryons. The 
color states could give rise to colorless hybrid baryons having 
singlet-singlet + octet-octet color components either of type 
$(qqq)(q \overline q)$  or  of type $(qqq)g$.  
In the context of a potential
model based on a one-gluon exchange hyperfine interaction, 
applied to the study of tetraquarks i. e. $q \overline q q \overline q$ systems, 
it has been shown \cite{WI,JMR,BRINK} that there is a
strong mixing between  states formed of two color singlet and two color octet
$(q \overline q)$ pairs which leads to a substantial lowering of the
variational energy.
This should also be the case for hybrid baryons
where singlet-singlet $ (qqq)_1 (q \overline q)_1$ and 
octet-octet $ (qqq)_8 (q \overline q)_8$ components could mix
substantially. The present study suggests that such hybrids are to be expected 
in all partial waves if only two-body $F_i \cdot F_j$ are considered.
This is in agreement with the findings of 
Ref. \cite{HELMINEN} based on a two-body confinement and
a schematic flavor and spin dependent hyperfine interaction
model. The subject deserves a separate investigation inasmuch as the
hybrid baryons raise a new interest \cite{PAGE}.

\vspace{1.5cm}
{\bf Acknowledgements.}
We are most grateful to Stephane Pepin and 
Jean-Marc Richard for
useful suggestions and a careful reading of the manuscript. 
We are also grateful to Gunnar Bali for useful correspondence
on lattice calculations.
One of us 
(Z.P.) acknowledges
financial support from the {\it Fond National de la Recherche
Scientifique} of Belgium and is grateful for warm hospitality
at the Theoretical Fundamental Physics Laboratory of the
University of Li\`ege. This work has been partially supported
by OTKA Grants T026233 and T029003.


\newpage
\begin{table} 
\caption{\label{octet}Three-quark color octet states
$[21]_C$ 
compatible with the Pauli principle. The first column gives the orbital
angular momentum $L$ and parity $P$, the second, third and fourth columns
give the permutation symmetry in the orbital, spin-isospin ad 
spin-isospin-color spaces respectively and the last column gives
an estimates of the excitation energy 
$E_N-E_0=(N+3) \hbar \omega [\overline \chi_8]^{1/2}
- 3 \hbar \omega$ of each state for a 
harmonic oscillator two-body confinement with $N$ quanta,
$\omega/c$ = 2 $fm^{-1}$ and $\overline \chi_8 = \chi_8/\chi_1$ with $\chi_8$ 
and $\chi_1$ from Eq. (\ref{CHI}) at $c = 0$. }
\begin{tabular}{cccccc}
\hline
$ L^P$ &  $[f]_O$ & $[f]_{IS}$ & $[f]_{ISC}$ & N & $E_N-E_0$\\
       &        &            &             &     &  (MeV)     \\
\hline  
$0^+$ & $[3]$    & $[21]$     & $[1^3]$     & 0 & 166 \\
$1^-$ & $[21]$   & $[3]$      & $[21]$      & 1 & 616 \\
$1^-$ & $[21]$   & $[21]$     & $[21]$      & 1 & 616 \\
$1^-$ & $[21]$   & $[1^3]$    & $[21]$      & 1 & 616 \\
$1^+$ & $[1^3]$   & $[21]$     & $[3]$       & 2 & 1066 \\ 
\end{tabular}          
\end{table}
\begin{table} 
\caption{\label{decuplet} Same as 
Table \ref{octet} but for color decuplet $[3]_C$ states
where $E_N-E_0=(N+3) \hbar \omega [\overline \chi_{10}]^{1/2}
- 3 \hbar \omega$ 
with $\overline \chi_{10} = \chi_{10}/\chi_1$ where $\chi_{10}$ and 
$\chi_1$ are from Eq. (\ref{CHI}) at $c = 0$. }
\begin{tabular}{cccccc}
\hline
$ L^P$ &  $[f]_O$ & $[f]_{IS}$ & $[f]_{ISC}$ & N & $E_N-E_0$\\
       &        &            &             &     &  (MeV)     \\      
\hline
$0^+$ & $[3]$    & $[1^3]$     & $[1^3]$     & 0  & 314 \\
$1^-$ & $[21]$   & $[21]$      & $[21]$      & 1  & 814 \\
$1^+$ & $[1^3]$  & $[3]$       & $[3]$       & 2  & 1312\\
\end{tabular}          
\end{table}
\begin{table}
\caption{\label{twobody} The expectation values 
of the two-body color operator $O_{ij}=\frac{7}{3} + F_i \cdot F_j$ 
of Eq. (\ref{V2b}) $(i<j)$ between three-quark states 
$|(c_2 c_3) [\tilde f]_C; c_1 [f]_C\rangle $
required in the Faddeev calculations. The particles 2 and 3 are first
coupled to a symmetric $[2]$ or an antisymmetric $[1^2]$ state and
next to particle 1 to a total color symmetry $[f]_C$.}
\begin{tabular}{cccc}
\hline
$[\tilde f]_C$ & $[f]_C$ & Color operator & Expectation value \\
\hline
$[2]$  & $[3]$   & $O_{ij}$ & 8/3 \\
$[2]$  & $[21]$  & $O_{23}$ & 8/3 \\
$[2]$  & $[21]$  & $O_{12}$ & 23/12 \\
$[2]$  & $[21]$  & $O_{13}$ & 23/12 \\
$[11]$ & $[21]$  & $O_{23}$ & 5/3 \\
$[11]$ & $[21]$  & $O_{12}$ & 29/12 \\
$[11]$ & $[21]$  & $O_{13}$ & 29/12 \\
$[11]$ & $[1^3]$ & $O_{ij}$ & 5/3 \\
\end{tabular}
\end{table} 
\begin{table}
\caption{\label{pauli} The quantum numbers 
of the components of a $q^3$ totally antisymmetric state
of color symmetry $[f]_C$, total angular momentum $L$, total spin S=1/2,
and total isospin I=1/2, compatible with the Pauli principle, 
as used in the Faddeev calculations.
Here $l$, $s$, $i$ and $[\tilde f]_C$ are the relative angular momentum, 
spin, isospin
and color quantum numbers of a two-quark pair and $\lambda$ is the relative
angular momentum of the third particle with respect to the pair.}
\begin{tabular}{ccccccc}
\hline
$[f]_C$  & $L$ & $l$ & $\lambda$ & $s$ & $i$ & $[\tilde f]_C$ \\
\hline
$[21]$ &   0 &   0   &   0   &   0   &   0   & $[11]$  \\
       &     &   0   &   0   &   1   &   1   & $[11]$  \\
       &     &   1   &   1   &   1   &   0   & $[11]$  \\
       &     &   1   &   1   &   0   &   1   & $[11]$  \\
       &     &   0   &   0   &   1   &   0   & $[2]$   \\
       &     &   0   &   0   &   0   &   1   & $[2]$   \\
       &     &   1   &   1   &   0   &   0   & $[2]$   \\
       &     &   1   &   1   &   1   &   1   & $[2]$   \\ 
\hline
$[21]$ &   1 &   0   &   1   &   0   &   0   & $[11]$  \\
       &     &   0   &   1   &   1   &   1   & $[11]$  \\
       &     &   1   &   0   &   0   &   1   & $[11]$  \\
       &     &   1   &   0   &   1   &   0   & $[11]$  \\
       &     &   1   &   0   &   0   &   0   & $[2]$  \\
       &     &   1   &   0   &   1   &   1   & $[2]$  \\
       &     &   0   &   1   &   0   &   1   & $[2]$  \\
       &     &   0   &   1   &   1   &   0   & $[2]$  \\  
\hline
$[3]$  &   0 &   0   &   0   &   1   &   0   & $[2]$  \\
       &     &   0   &   0   &   0   &   1   & $[2]$  \\
       &     &   1   &   1   &   0   &   0   & $[2]$  \\  
       &     &   1   &   1   &   1   &   1   & $[2]$  \\      
\end{tabular}
\end{table}

\newpage

\newpage

\newpage

\end{document}